\documentclass[12pt,epsfig]{article}
\usepackage[xdvi]{graphicx}

\newcommand{\beq}{\begin{equation}}
\newcommand{\eeq}{\end{equation}}
\newcommand{\bea}{\begin{eqnarray}}
\newcommand{\eea}{\end{eqnarray}}

\def\e2sig{e^{-2r\sigma}}

\begin{document}
\setlength{\baselineskip}{0.7cm}
\begin{titlepage} 
\begin{flushright}
KOBE-TH-07-11 \\
KEK-TH-1205 
\end{flushright}
\vspace*{10mm}
\begin{center}{\LARGE\bf Gauge-Higgs Unification at LHC}
\end{center}
\vspace*{10mm}
\begin{center}
{\Large Nobuhito Maru}$^{a,}$\footnote{E-mail: maru@people.kobe-u.ac.jp} and 
{\Large Nobuchika Okada}$^{b,}$\footnote{E-mail: okadan@post.kek.jp}
\end{center}
\vspace*{0.2cm}
\begin{center}
${}^{a}${\it Department of Physics, Kobe University, 
Kobe 657-8501, Japan}
\\[0.2cm]
${}^{b}${\it Theory Division, KEK, 
1-1 Oho, Tsukuba 305-0801, Japan} 
\end{center}
\vspace*{2cm}
\begin{abstract}
Higgs boson production by the gluon fusion and 
 its decay into two photons at the LHC 
 are investigated in the context of the gauge-Higgs unification scenario. 
The qualitative behaviors for these processes 
 in the gauge-Higgs unification 
 are quite distinguishable from those of the Standard Model 
 and the universal extra dimension scenario 
 because of the overall sign difference for the effective couplings 
 induced by one-loop corrections through Kaluza-Klein (KK) modes. 
For the KK mode mass smaller than 1 TeV, 
 the Higgs production cross section and its branching ratio 
 into two photons are sizably deviated from those 
 in the Standard Model. 
Associated with the discovery of Higgs boson, 
 this deviation may be measured at the LHC. 
\end{abstract}
\end{titlepage}

\section{Introduction}
The gauge-Higgs unification \cite{Manton} is a very fascinating 
 scenario beyond the Standard Model (SM) 
 since the SM Higgs doublet is identified with the extra component of 
 the higher dimensional gauge field 
 and its mass squared correction is predicted to be finite \cite{HIL} 
 regardless of the non-renormalizable theory. 
This fact has opened a new possibility to solve 
 the gauge hierarchy problem without, for example, supersymmetry. 
The finiteness of Higgs mass has been discussed and checked 
 by the various explicit calculations \cite{ABQ}. 
Furthermore there has attracted a large amount of attention 
 from the various viewpoints \cite{Hosotani}-\cite{PW}.

The Large Hadron Collider (LHC) will start its operation soon 
 and the collider signatures of various new physics models 
 beyond the SM have been extensively studied. 
However, as far as we know, 
 the gauge-Higgs unification has not been 
 so much explored from this respect. 
The gauge-Higgs unification shares 
 the similar structure with the universal extra dimension (UED) 
 scenario \cite{UED-org1} 
 \cite{UED-org2}, namely, 
 in effective four dimensional theory, 
 Kaluza-Klein (KK) states of the Standard Model particles appear. 
The collider phenomenology on the KK particles will be 
 quite similar to the one in the UED scenario. 
A crucial difference should lie in the Higgs sector, 
 because the Higgs doublet originates from the higher 
 dimensional gauge field. 
The discovery of Higgs boson is expected at the LHC, by which 
 the origin of the electroweak symmetry breaking 
 and the mechanism responsible for generating fermion masses 
 will be revealed. 
Precise measurements of Higgs boson properties will 
 provide us the information of a new physics relevant 
 to the Higgs sector.

In this paper, we investigate the effect of gauge-Higgs unification 
 on Higgs boson phenomenology at the LHC, 
 namely, the production and decay processes of Higgs boson. 
At the LHC, the gluon fusion is the dominant Higgs boson production
 process and for light Higgs boson with mass $m_h < 150$ GeV, 
 two photon decay mode of Higgs boson becomes 
 the primary discovery mode \cite{Djouadi} 
 nevertheless its branching ratio is ${\cal O}(10^{-3})$. 
The coupling between Higgs boson and these gauge bosons 
 are induced through quantum corrections at one-loop level 
 even in the Standard Model. 
Therefore, we can expect a sizable effect 
 from new particles if they contribute to the coupling 
 at one-loop level. 
In a five dimensional gauge-Higgs unification model, 
 we calculate one-loop diagrams with KK fermions 
 for the effective couplings between Higgs boson and the gauge bosons 
 (gluons and photons). 
If the KK mass scale is small enough, 
 we can see a sizable deviation from the SM couplings 
 and as a result, the number of signal events from Higgs production 
 at the LHC can be altered from the SM one. 
Interestingly, reflecting the special structure of Higgs sector 
 in the gauge-Higgs unification, 
 there is a clear qualitative difference from the UED scenario, 
 the signs of the effective couplings are opposite 
 to those in the UED scenario .

\section{Toy Model}
In this paper, we consider a toy model 
 of five dimensional (5D) $SU(3)$ gauge-Higgs unification  
 with an orbifold $S^1/Z_2$ compactification, 
 in order to avoid unnecessary complications for our discussion. 
Although the predicted Weinberg angle in this toy model 
 is unrealistic, $\sin^{2} \theta_{W} = \frac{3}{4}$, 
 this does not affect our analysis. 
We introduce an $SU(3)$ triplet fermion as a matter field, 
 which is identified with top and bottom quarks 
 and their KK excited states, 
 although the top quark mass vanishes and the bottom quark mass 
 $m_{b} = M_{W}$ in this toy model. 
In this work we neglect other generations 
 since the effects of light generations are very small 
 comparing to the effect by the top quark.

The $SU(3)$ gauge symmetry is broken to $SU(2) \times U(1)$ 
 by the orbifolding on $S^1/Z_2$ 
 and adopting a non-trivial $Z_2$ parity assignment 
 for the members of an irreducible representation of $SU(3)$, 
 as stated below. 
The remaining gauge symmetry $SU(2) \times U(1)$ is supposed to be broken 
 by the vacuum expectation value (VEV) of the zero-mode of $A_5$, 
 the extra space component of the gauge field identified with 
 the SM Higgs doublet, through the Hosotani mechanism \cite{Hosotani}. 
We do not address the origin of $SU(2) \times U(1)$ gauge symmetry 
 breaking and the resultant Higgs boson mass 
 in the one-loop effective Higgs potential, 
 which is highly model-dependent and out of our scope in this paper.

The Lagrangian is simply given by 
\bea
{\cal L} = -\frac{1}{2} \mbox{Tr}  (F_{MN}F^{MN}) 
+ i\bar{\Psi}D\!\!\!\!/ \Psi
\label{lagrangian}
\eea
where $\Gamma^M=(\gamma^\mu, i \gamma^5)$, 
\bea
F_{MN} &=& \partial_M A_N - \partial_N A_M -i g_{5} [A_M, A_N]~(M,N = 0,1,2,3,5), \\
D\!\!\!\!/ &=& \Gamma^M (\partial_M -ig_{5} A_M) \ \ 
(A_{M} = A_{M}^{a} \frac{\lambda^{a}}{2} \ 
(\lambda^{a}: \mbox{Gell-Mann matrices})),  \\
\Psi &=& (\psi_1, \psi_2, \psi_3)^T.
\eea
The periodic boundary conditions are imposed along $S^1$ for all fields. 
The non-trivial $Z_2$ parities are assigned for each field  as follows, 
\bea 
\label{z2parity} 
A_\mu = 
\left(
\begin{array}{ccc}
(+,+) & (+,+) & (-,-) \\
(+,+) & (+,+) & (-,-) \\
(-,-) & (-,-) & (+,+) 
\end{array}
\right), \ \ 
A_5 = 
\left(
\begin{array}{ccc}
(-,-) & (-,-) & (+,+) \\
(-,-) & (-,-) & (+,+) \\
(+,+) &(+,+) & (-,-)
\end{array}
\right), 
\eea
\bea 
\label{fermizero} 
\Psi = 
\left(
\begin{array}{cc}
\psi_{1L}(+,+) + \psi_{1R}(-, -) \\
\psi_{2L}(+,+) + \psi_{2R}(-, -) \\
\psi_{3L}(-,-) + \psi_{3R}(+, +) \\
\end{array}
\right),
\eea
where $(+,+)$ means that $Z_2$ parities are even at the fixed points $y=0$ 
and $y = \pi R$, for instance. $y$ is the fifth coordinate and 
$R$ is the compactification radius. 
$\psi_{1L} \equiv \frac{1}{2}(1-\gamma_5)\psi_1$, etc.

Following these boundary conditions, 
 KK mode expansions for the gauge fields 
 and the fermions are carried out.   
\bea
A_{\mu,5}^{(+,+)}(x,y) &=& \frac{1}{\sqrt{2 \pi R}} 
\left[
A_{\mu,5}^{(0)}(x) + \sqrt{2} \sum_{n=1}^\infty A_{\mu,5}^{(n)}(x) 
\cos (ny/R)
\right], \\
A_{\mu,5}^{(-,-)}(x,y) &=& \frac{1}{\sqrt{\pi R}} 
\sum_{n=1}^\infty A_{\mu,5}^{(n)}(x) 
\sin (ny/R), \\
\psi_{1L, 2L, 3R}^{(+,+)}(x,y) &=& \frac{1}{\sqrt{2 \pi R}} 
\left[
\psi_{1L, 2L, 3R}^{(0)}(x) 
+ \sqrt{2} \sum_{n=1}^\infty \psi_{1L,2L,3R}^{(n)}(x) \cos (ny/R)
\right], \\
\psi_{3L,1R,2R}^{(-,-)}(x,y) &=& i \ \frac{1}{\sqrt{\pi R}} 
\sum_{n=1}^\infty \psi_{3L,1R,2R}^{(n)}(x) \sin (ny/R). 
\label{KK}
\eea 
For the zero-mode of bosonic sector, 
 we obtain exactly what we need for the Standard Model: 
\bea
A^{(0)}_{\mu} = \frac{1}{2}
\left(
\begin{array}{ccc}
W^{3}_{\mu}+ \frac{B_{\mu}}{\sqrt{3}} & \sqrt{2} W^{+}_{\mu} & 0 \\
\sqrt{2} W^{-}_{\mu} & - W^{3}_{\mu}+ \frac{B_{\mu}}{\sqrt{3}} & 0 \\
0& 0 & -\frac{2}{\sqrt{3}} B_{\mu}   
\end{array}
\right) , \ \ 
A_5^{(0)} = \frac{1}{\sqrt{2}}
\left(
\begin{array}{ccc}
0 & 0 & h^{+} \\
0 & 0 & h^{0} \\
h^{-} & h^{0\ast} & 0 
\end{array}
\right),   
\eea
where $W_\mu^{3}, \ W_{\mu}^{\pm}$, $B_\mu$ 
 are $h = (h^{+}, h^{0})^{t}$ is the Higgs doublet. 
For the zero mode in the fermion sector, 
 a fermion corresponding to the right-handed top quark 
 $t_{R}$ is missing as we mentioned above, 
\bea
\Psi^{(0)} = 
\left(
\begin{array}{cc}
t_{L} \\
b_{L} \\
b_{R} \\
\end{array}
\right). 
\eea
In order to obtain a realistic model, 
 more elaborate gauge-Higgs unification model 
 should be considered. 
The $SU(2)_{L} \times U(1)_{Y}$ gauge symmetry is broken 
 by the Higgs VEV, $\langle h^{0} \rangle = v/\sqrt{2}$, 
 in other words, $ \langle A_{5} \rangle = v/2 \; \lambda_{6}$.

After the gauge symmetry breaking, 4D effective Lagrangian 
 among KK fermions, the SM gauge boson and Higgs boson ($h$) 
 defined as $h^0=(v+h)/\sqrt{2}$ can be derived from the term 
 ${\cal L}_{{\rm fermion}}  = i\bar{\Psi}D\!\!\!\!/ \Psi$ 
 in Eq.~(\ref{lagrangian}). 
Integrating over the fifth dimensional coordinate, 
 we obtain a 4D effective Lagrangian:  
\bea
{\cal L}_{{\rm fermion}}^{(4D)} 
&=& 
\sum_{n=1}^{\infty} \left\{  
i(\bar{\psi}_1^{(n)}, \bar{\psi}_2^{(n)}, \bar{\psi}_3^{(n)}) 
\gamma^\mu \partial_\mu  
\left(
\begin{array}{c}
\psi_1^{(n)} \\
\psi_2^{(n)} \\
\psi_3^{(n)}
\end{array}
\right) \right. \nonumber \\ 
&& \left. 
+\frac{g}{2} (\bar{\psi}_1^{(n)}, \bar{\psi}_2^{(n)}, \bar{\psi}_3^{(n)})  
\left(
\begin{array}{ccc}
W^{3}_{\mu}+ \frac{B_{\mu}}{\sqrt{3}} & \sqrt{2} W^{+}_{\mu} & 0 \\
\sqrt{2} W^{-}_{\mu} & - W^{3}_{\mu}+ \frac{B_{\mu}}{\sqrt{3}} & 0 \\
0& 0 & -\frac{2}{\sqrt{3}} B_{\mu}   
\end{array}
\right) 
\gamma^{\mu} 
\left(
\begin{array}{c}
\psi_1^{(n)} \\
\psi_2^{(n)} \\
\psi_3^{(n)}
\end{array}
\right) \right. \nonumber  \\ 
&& \left. 
- (\bar{\psi}_1^{(n)}, \bar{\psi}_2^{(n)}, \bar{\psi}_3^{(n)})    
\left(
\begin{array}{ccc}
m_{n} & 0 & 0 \\
0 & m_{n} & -(m+gh) \\
0& -(m+gh)  & m_{n} 
\end{array}
\right)
\left(
\begin{array}{c} 
\psi_1^{(n)} \\
\psi_2^{(n)} \\
\psi_3^{(n)}
\end{array}
\right) \right\} \nonumber \\ 
&&  + i \bar{t}_{L} \gamma^{\mu} \partial_{\mu} t_{L} 
+  \bar{b} (i \gamma^{\mu} \partial_{\mu} - m -g h) b 
\nonumber \\ 
&& +\frac{g}{\sqrt{2}} (\bar{t}\gamma_{\mu} P_L b  W^{+\mu} 
+ \bar{b}\gamma_{\mu} P_L t W^{-\mu}) 
+ \frac{g}{2} (\bar{t}\gamma_{\mu} P_L t 
- \bar{b}\gamma_{\mu} P_L b) W_{3}^{\mu} \nonumber \\ 
&& + \frac{\sqrt{3}g}{6} (\bar{t}\gamma_{\mu} P_L t 
+ \bar{b}\gamma_{\mu} P_L b -2\bar{b}\gamma_{\mu} P_R b) B^{\mu}, 
\label{4Deffaction}
\eea
where $P_L \equiv \frac{1}{2}(1 -\gamma_5)$, $m_{n} = \frac{n}{R}$, 
 $g = \frac{g_{5}}{\sqrt{2\pi R}}$ is the 4D gauge coupling, 
 and $m = \frac{gv}{2} (= M_{W})$ is the bottom quark mass 
 in this toy model. 
In deriving the 4D effective Lagrangian (\ref{4Deffaction}), 
 a chiral rotation 
\beq  
\psi_{1,2,3} \ \to \ e^{-i\frac{\pi}{4}\gamma_{5}} \psi_{1,2,3}  
\eeq
 has been made in order to get rid of $i \gamma_{5}$.

We easily see that the mass matrix for the KK modes 
can be diagonalized by use of the mass eigenstates 
$\tilde{\psi}_{2}^{(n)}, \ \tilde{\psi}_{3}^{(n)}$,   
\bea 
\pmatrix{ 
\psi_{1}^{(n)} \cr 
\tilde{\psi}_{2}^{(n)} \cr 
\tilde{\psi}_{3}^{(n)} \cr 
} 
= U 
\pmatrix{ 
\psi_{1}^{(n)} \cr 
\psi_{2}^{(n)} \cr 
\psi_{3}^{(n)} \cr 
}, \ \ \  
U =\frac{1}{\sqrt{2}}
\left(
\begin{array}{ccc}
\sqrt{2} & 0 & 0 \\
0 & 1 & -1 \\
0 & 1 & 1 
\end{array}
\right), 
\eea
as 
\bea 
U \ 
\left(
\begin{array}{ccc}
m_{n} & 0 & 0 \\
0 & m_{n} & -m  \\
0& -m  & m_{n} 
\end{array}
\right) 
\ U^{\dagger} 
= 
\left(
\begin{array}{ccc}
m_n & 0 & 0 \\
0 & m_n + m & 0 \\
0 & 0 & m_n-m 
\end{array}
\right). 
\eea
Note that the mass splitting $m_\pm^{(n)} \equiv m_{n} \pm m$ 
 occurs associated with a mixing 
 between the $SU(2)$ doublet component and singlet component. 
Each of mass eigenvalues has a periodicity with respect to $m$: 
 $m_{n} \pm (m + \frac{1}{R}) = m_{n \pm 1} \pm m$, 
 which is a remarkable feature of the gauge-Higgs unification, 
 not shared in the UED scenario, 
 where the mass of KK modes are given by $\sqrt{m_{n}^{2} + m^{2}}$.

In terms of the mass-eigenstates for non-zero KK modes, 
 the Lagrangian is described as 
\bea
&&{\cal L}_{{\rm fermion}}^{(4D)}  
= \sum_{n=1}^{\infty} \left\{  (\bar{\psi}_1^{(n)}, 
\bar{\tilde{\psi}}_2^{(n)}, \bar{\tilde{\psi}}_3^{(n)}) 
\right. \nonumber \\ 
&& \left. \times \left(
\begin{array}{ccc}
i \gamma^{\mu} \partial_{\mu} - m_{n} & 0 & 0 \\
0 & i \gamma^{\mu} \partial_{\mu} 
 -\left( m_{+}^{(n)} + \frac{m}{v} h \right) & 0 \\
0& 0 &i \gamma^{\mu} \partial_{\mu} 
 -\left( m_{-}^{(n)} - \frac{m}{v} h \right)  
\end{array}
\right)
\left(
\begin{array}{c} 
\psi_1^{(n)} \\
\tilde{\psi}_2^{(n)} \\
\tilde{\psi}_3^{(n)}
\end{array}
\right) \right. \nonumber \\ 
&& \left. 
+\frac{g}{2} (\bar{\psi}_1^{(n)}, \bar{\tilde{\psi}}_2^{(n)}, 
\bar{\tilde{\psi}}_3^{(n)})  
\left(
\begin{array}{ccc}
W^{3}_{\mu}+ \frac{\sqrt{3}B_{\mu}}{3} & W^{+}_{\mu} & W^{+}_{\mu} \\
W^{-}_{\mu} & - \frac{W^{3}_{\mu}}{2} - \frac{\sqrt{3} B_{\mu}}{6} & 
- \frac{W^{3}_{\mu}}{2} + \frac{\sqrt{3} B_{\mu}}{2}  \\
W^{-}_{\mu} & - \frac{W^{3}_{\mu}}{2} + \frac{\sqrt{3} B_{\mu}}{2} & 
- \frac{W^{3}_{\mu}}{2} - \frac{\sqrt{3} B_{\mu}}{6} 
\end{array}
\right) 
\gamma^{\mu} 
\left(
\begin{array}{c}
\psi_1^{(n)} \\
\tilde{\psi}_2^{(n)} \\
\tilde{\psi}_3^{(n)}
\end{array}
\right) \right\} \nonumber  \\ 
&&   + \ \mbox{zero-mode part}.  
\label{4Deff}
\eea
The relevant Feynman rules for our calculation can be 
 read off from this Lagrangian.  
Note that the mass eigenstate for $m_{+}^{(n)}$ 
 has the Yukawa coupling $-m/v$, which is exactly the same 
 as the one for the zero mode, 
 while the Yukawa coupling of the mass eigenstate for $m_{-}^{(n)}$ 
 has an opposite sign, $+m/v$. 
Together with the mass splitting of KK modes, 
 this property is a general one realized in any 
 gauge-Higgs unification model 
 and leads to a clear qualitative difference 
 of the gauge-Higgs unification from the UED scenario, 
 as we will see.

\section{Effective couplings between Higgs boson and gauge bosons}

Before calculating contributions of KK fermions 
 to one-loop effective couplings between Higgs boson 
 and gauge bosons (gluons and photons), 
 it is instructive to recall the SM result. 
We parameterize the effective coupling between 
 Higgs boson and gluons as 
\bea 
{\cal L}_{\rm eff} = C_g^{SM} \;  h \; G^{a\mu\nu}G^a_{\mu\nu}, 
\eea
 where $G_{\mu\nu}^a$ is a gluon field strength tensor. 
This coupling is generated by one-loop corrections 
 (triangle diagram) on which quarks are running. 
The top quark loop diagram gives the dominant contribution 
 and the coupling $C_g^{SM}$ is described in 
 the following instructive form: 
\bea 
 C_g^{SM} = - \frac{m_t}{v} \times 
          \frac{\alpha_s F_{1/2}(4m_t^2/m_h^2)}{8\pi m_t} 
           \times \frac{1}{2},  
\eea 
where, in the right hand side, 
 the first term, $- \frac{m_t}{v}$, is top Yukawa coupling, 
 the second term is from the loop integral 
 with the QCD coupling $\alpha_s$ at QCD vertexes, 
 the loop function $F_{1/2}(\tau)$ given by (for $\tau \geq 1$) 
\bea
 F_{1/2}(\tau) &=& -2 \tau \left( 
   1+ \left( 1 - \tau \right) 
   [\sin^{-1}(1/\sqrt{\tau})]^2 \right) \nonumber \\
   &\to& -\frac{4}{3}~~{\rm for}~\tau \gg 1 ,  
\label{loopfunc}
\eea
 and $1/2$ is a QCD group factor (Dynkin index). 
Mass of the fermion (top quark) running the loop 
 appears in the denominator in the second term, 
 which is canceled with top quark mass from Yukawa coupling. 
It is well-known that in the top quark decoupling limit, 
 namely top quark mass $m_t$ is much heavier 
 than Higgs boson mass $m_h$, 
 $F_{1/2}$  becomes a constant and 
 the resultant effective coupling becomes independent 
 of $m_t$ and $m_h$.

Calculations of KK mode contributions are completely analogous 
 to the top loop correction. 
The structure described in our toy model is common 
 in any gauge-Higgs unification model 
 and in some realistic model, 
 we will have KK modes of top quark 
 with mass eigenvalue $m_{\pm}^{(n)}= m_n \pm m_t$ 
 and Yukawa couplings $ \mp m_t/v$, respectively. 
The KK mode contributions are found to be  
\bea
{\cal L}_{{\rm eff}} &=& C_g^{KK(GH)} \; 
  h \; G^{a\mu\nu} G^a_{\mu\nu} , \nonumber \\ 
 C_g^{KK(GH)} &=& -\sum_{n=1}^\infty 
 \left[
 \frac{m_t}{v} \times 
 \frac{\alpha_s F_{1/2}(4 m_{+}^{(n)2}/m_h^2)}
 {8\pi m_{+}^{(n)}} \times \frac{1}{2} \right]
+ \sum_{n=1}^\infty 
 \left[
 \frac{m_t}{v} \times 
 \frac{\alpha_s F_{1/2}(4 m_{-}^{(n)2}/m_h^2)}
 {8\pi m_{-}^{(n)}} \times \frac{1}{2} \right] \nonumber \\ 
&\simeq & 
 \frac{m_t \alpha_s}{12 \pi v}
 \sum_{n=1}^\infty 
\left[ \frac{1}{m_{+}^{(n)}} - \frac{1}{m_{-}^{(n)}} \right] 
 \simeq  - \frac{\alpha_s}{6 \pi v}
 \sum_{n=1}^\infty \frac{m_t^2}{m_n^2}  
\eea
where we have taken the limit $m_h^2,~m_t^2 \ll m_n^2$, 
 to simplify the results. 
Note that this result is finite and 
 this finiteness is a consequence of cancellation 
 between two divergent corrections with opposite signs. 
Also, note that the KK mode contribution is subtractive 
 against the top quark contribution in the SM.

It is interesting to compare our result to 
 that in the UED scenario \cite{UED, UED2}, 
 where the KK mode mass spectrum and Yukawa couplings 
 are given by 
 $M_n = \sqrt{m_n^2 + m_t^2}$ without mass splitting 
 and $-(m_t/v) \times (m_t/M_n)$, respectively. 
In this case, we find 
\bea 
{\cal L}_{{\rm eff}} &=& C_g^{KK(UED)} \; 
  h \; G^{a\mu\nu} G^a_{\mu\nu} , \nonumber \\ 
 C_g^{KK(UED)} &=& 
 -\sum_{n=1}^\infty 
 \left[
 \frac{m_t}{v} \frac{m_t}{M_n}
 \times 
 \frac{\alpha_s F_{1/2}(4 M_n^2/m_h^2)}{8 \pi M_n} 
 \times \frac{1}{2} \right]  \times 2 \nonumber \\ 
&\simeq & 
 \frac{\alpha_s}{6 \pi v}
  \sum_{n=1}^\infty \frac{m_t^2}{m_n^2}  
\label{gghUED}
\eea
where we have, again, taken 
 the limit $m_h^2,~m_t^2 \ll m_n^2$, to simplify the result. 
In the limit, we arrive at the same result as the one 
 in the gauge-Higgs unification model, except for the sign. 
This KK mode contribution is constructive to 
 the top quark one in the SM.

The contribution of top quark KK modes 
 to the effective coupling between Higgs boson and photons 
 are calculated in the same way. 
In fact, the final result can be obtained by  
 the replacements, 
 $\alpha_s \to \alpha_{em}$ 
 and the group factor $1/2 \to Q_t^2 \times 3 $, 
 top quark electric charge$^2 \times $number of colors:  
\bea
{\cal L}_{{\rm eff}} &=& C_{\rm \gamma}^{KK(GH)} \; 
  h \; F^{\mu\nu} F_{\mu\nu} , \nonumber \\ 
 C_{{\rm \gamma}}^{KK(GH)} &=& -\sum_{n=1}^\infty 
 \left[
 \frac{m_t}{v} \times 
 \frac{\alpha_{em} F_{1/2}(4 m_{+}^{(n)2}/m_h^2)}
 {8\pi m_{+}^{(n)}} \times \frac{4}{3} \right]
+ \sum_{n=1}^\infty 
 \left[
 \frac{m_t}{v} \times 
 \frac{\alpha_{em} F_{1/2}(4 m_{-}^{(n)2}/m_h^2)}
 {8\pi m_{-}^{(n)}} \times \frac{4}{3} \right] \nonumber \\ 
&\simeq & 
 \frac{2 m_t \alpha_{em}}{9 \pi v}
 \sum_{n=1}^\infty 
\left[ \frac{1}{m_{+}^{(n)}} - \frac{1}{m_{-}^{(n)}} \right] 
 \simeq  - \frac{4 \alpha_{em}}{9 \pi v}
 \sum_{n=1}^\infty \frac{m_t^2}{m_n^2} ,  
\label{gghGH}
\eea
where we have taken the limit $m_h^2,~m_t^2 \ll m_n^2$, 
 to simplify the results. 
For the effective coupling with photons, 
 in addition to the KK fermion contributions, 
 there is another contribution, namely 
 the KK W-boson loop corrections, as in the SM. 
This calculation is quite complicated, because we have to 
 include contributions by KK Nambu-Goldstone bosons and KK ghosts, 
 according to the five dimensional gauge invariance. 
In this paper, we neglect such contributions 
 compared to those from the KK top quark contributions 
 in the following reasons: 
First, the KK mode contributions are decoupling effects, 
 and the KK top quark and KK W-boson loop contributions 
 are proportional to top quark mass squared and W-boson mass squared, 
 respectively. 
Top quark is much heavier than W-boson, 
 so that KK top quark contributions are likely to be dominant. 
Second, in the gauge-Higgs unification, 
 Yukawa coupling is nothing but the gauge coupling
 and a fermion mass is naturally the same as W-boson mass 
 and this is too small for the realistic top quark mass. 
One way to ameliorate this problem 
 is to introduce a large dimensional representation 
 as discussed in \cite{CCP}, in which the SM top quark is implemented, 
 so that top Yukawa coupling can be correctly reproduced 
 with a Clebsch-Gordan coefficient (a factor 2 is suitable). 
In this setup, the effective 4D theory includes 
 extra vector-like top-like quarks and its KK modes. 
Thus, fermion KK mode contributions are enhanced by 
 a number of extra top-like quarks. 
Third, in some gauge-Higgs unification models, 
 bulk top-like quarks with the half-periodic boundary condition 
 are often introduced to realize the correct electroweak symmetry 
 breaking and a Higgs boson mass consistent 
 with the current experimental lower bound. 
The lowest KK mass of the half-periodic fermions 
 is half of the lowest KK mass of periodic fields, 
 so that their loop contributions can dominate over 
 those by periodic KK mode fields. 

In the SM, both the top and W-boson loop corrections should be  
 taken into account, because of non-decoupling effects 
 that for a light Higgs boson, the effective coupling 
 is not so sensitive to top and W-boson masses. 
The effective coupling between Higgs boson and two photons 
 is given by 
\bea
 {\cal L}_{{\rm eff}} = C_{\gamma}^{SM} h F^{\mu\nu} F_{\mu\nu}, 
\eea 
where the coupling is the sum of 
 top loop contribution ($C_{\gamma,t}^{SM}$) and 
 W-boson loop contribution ($C_{\gamma,W}^{SM}$) such as 
\bea 
 C_{\gamma, t}^{SM} &=& 
 -\frac{m_t}{v} \times  
 \frac{\alpha_{em} F_{1/2}(4m_t^2/m_h^2)}{8\pi m_t} 
 \times \frac{4}{3},  \nonumber \\
 C_{\gamma, W}^{SM} &=& 
  - \frac{m_W^2}{v} \times 
   \frac{\alpha_{em} F_1(4m_W^2/m_h^2)}{8 \pi m_W^2}  
\eea
with the loop function,   
\bea 
 F_1(\tau) &=& 
 2 + 3 \tau +3 \tau (2-\tau) [\sin^{-1}(1/\sqrt{\tau})]^2 
 \nonumber \\
&\rightarrow & 7 \; \mbox{for} \; \tau \gg 1. 
\eea 
In the SM, signs of the top quark and W-boson loop contributions 
 are opposite to each other and the W-loop contribution dominates 
 for the effective coupling. 
Therefore, the fermion KK mode contributions in the gauge-Higgs
 unification model is constructive to the SM one.

\section{Effects on Higgs boson search at LHC} 
As we have shown, the KK mode loop contribution to 
 the effective coupling between Higgs boson and 
 gluons or photons is subtractive to 
 the top quark loop contribution in the SM. 
This fact leads to remarkable effects on Higgs boson 
 search at the LHC.  
Since the main production process of Higgs boson at the LHC 
 is through gluon fusion, so that the deviation 
 of the effective coupling between Higgs boson and gluons 
 directly affects the Higgs boson production cross section. 
When Higgs boson is light $m_h < 150$ GeV, 
 the primary discovery mode of Higgs boson is 
 its two photon decay channel. 
Therefore, the deviation of the effective coupling 
 between Higgs boson and photons from the SM one 
 gives important effect on the number of two photon events 
 from Higgs boson decay.

\begin{figure}[t]
\includegraphics[scale=1.2]{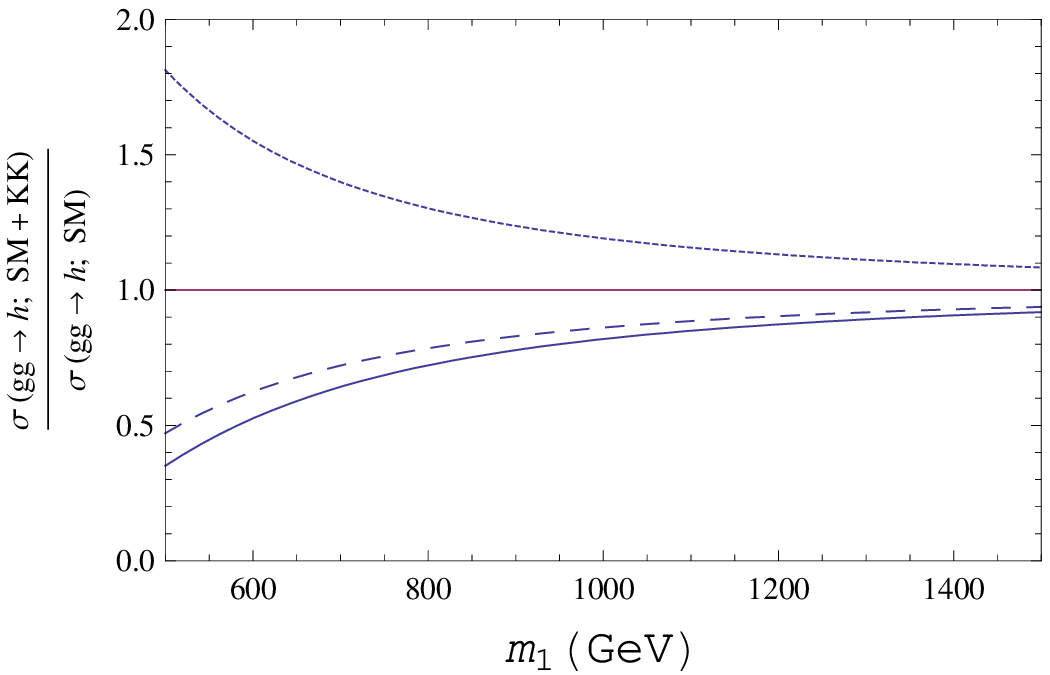}
\caption{
The ratio of the Higgs boson production cross sections 
 in the gauge-Higgs unification scenario and in the SM, 
 as a function of the KK mode mass $m_1$. 
The solid and dashes lines corresponds to 
 the results including the periodic and the half-periodic 
 fermion contributions, respectively. 
As a reference, the result in the UED scenario 
 with top quark KK modes is also shown (dotted line). 
Here (in all Figures), we have taken $m_h=120$ GeV. 
}
\label{fig1}
\end{figure}

Let us first consider the ratio of the Higgs boson production
 cross section in the gauge-Higgs unification model to the SM one, 
 which is described as 
\bea 
  \frac{\sigma(gg \to h;~{\rm SM+KK})}
       {\sigma(gg \to h;~{\rm SM})}  =
  \left( 1 +  \frac{C_g^{KK(GH)}}{C_g^{SM}} \right)^2 . 
\eea
The results are depicted in Fig.~\ref{fig1} 
 as a function of the mass of the lightest KK mode (diagonal) 
 mass eigenvalue ($m_1$). 
For the bulk fermion with the periodic boundary condition, 
 $m_1 = 1/R$ with the fifth dimensional radius $R$, 
 while we define $m_1=1/(2 R)$ for the bulk fermion 
 with the half-periodic boundary condition. 
In this analysis, we take $m_h=120$ GeV.  
The result is not sensitive to the Higgs boson mass 
 if $m_h < 2 m_t$. 
For reference, the result in the UED scenario is also shown, 
 for which only the periodic fermion has been considered. 
The KK fermion contribution is subtractive 
 and the Higgs production cross section is reduced 
 in the gauge-Higgs unification scenario, 
 while it is increased in the UED scenario. 
This is a crucial point to distinguish 
 the gauge-Higgs unification from the UED scenario. 
Interestingly, even for $m_1=1$ TeV, the KK fermion contribution 
 is sizable and the production cross section is reduced by 
 about 18\%.

\begin{figure}[t]
\includegraphics[scale=1.2]{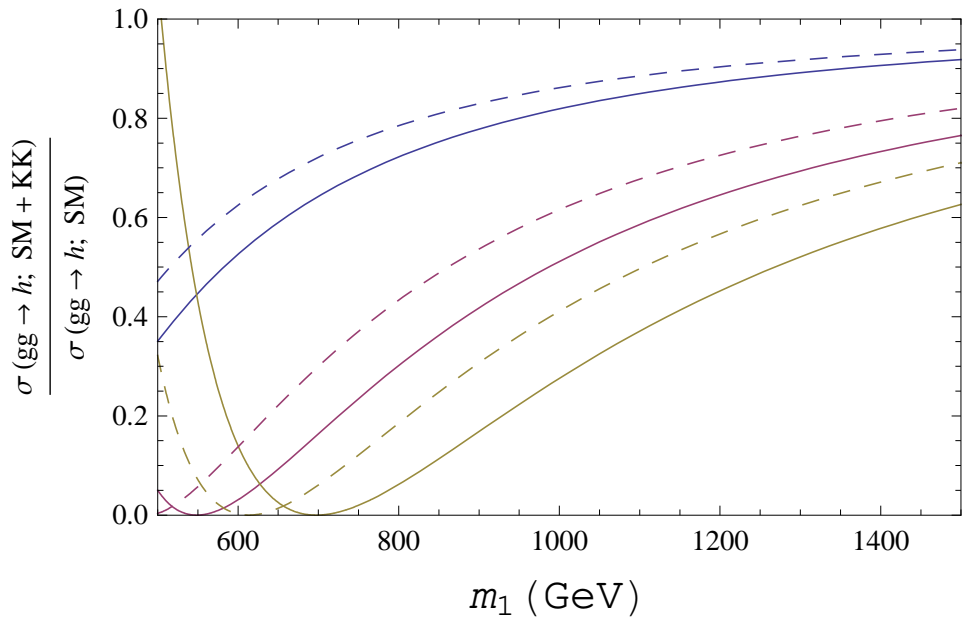}
\caption{
The ratio of the Higgs boson production cross sections 
 in the gauge-Higgs unification scenario 
 with $n_t$ periodic and half-periodic KK modes 
 and in the SM, as a function of the KK mode mass $m_1$. 
The solid lines represent the results 
 including the periodic KK fermion contributions. 
Each solid line corresponds  to 
 $n_t=$1, 3 and 5 from top to bottom at $m_1=1500$ GeV. 
The results for the $n_t$ half-periodic fermions 
 are depicted as the dashed lines, 
 corresponding to $n_t=$1, 3 and 5 
 from top to bottom at $m_1=1500$ GeV. 
The results for $n_t=1$ are those shown in Fig.~\ref{fig1}. 
}
\label{fig2}
\end{figure}

As mentioned before, in a realistic model, 
 top quark would be implemented in a large representation fermion. 
If this is the case, the effective 4D theory includes 
 extra top-like quarks and their KK modes. 
If such extra top-like quarks appear, the effective Higgs boson 
 coupling receives more contributions. 
In Fig.~\ref{fig2}, we show the ratio 
 in the case that $n_t$ KK towers of top-like quark multiplets 
 are introduced. 
In this case, the KK mode contributions are enhanced by 
 the replacement $C_g^{KK(GH)} \to C_g^{KK(GH)} \times n_t$  
 ($n_t=1$ corresponds to Fig.~\ref{fig1}). 
The value of $n_t$ is highly model-dependent. 
As $n_t$ becomes large, the KK mode contributions can even dominate 
 over the effective coupling of the SM. 
In other words, the Higgs boson production cross section 
 can be quite altered in the gauge-Higgs unification scenario. 
This happens also in the UED scenario, if extra top-like 
 fermions are introduced. 
However, in the UED scenario, there is no constraint (or prediction) 
 in the Yukawa and Higgs sectors and there is no positive motivation 
 for introduction of extra fermions.

\begin{figure}[t]
\includegraphics[scale=1.2]{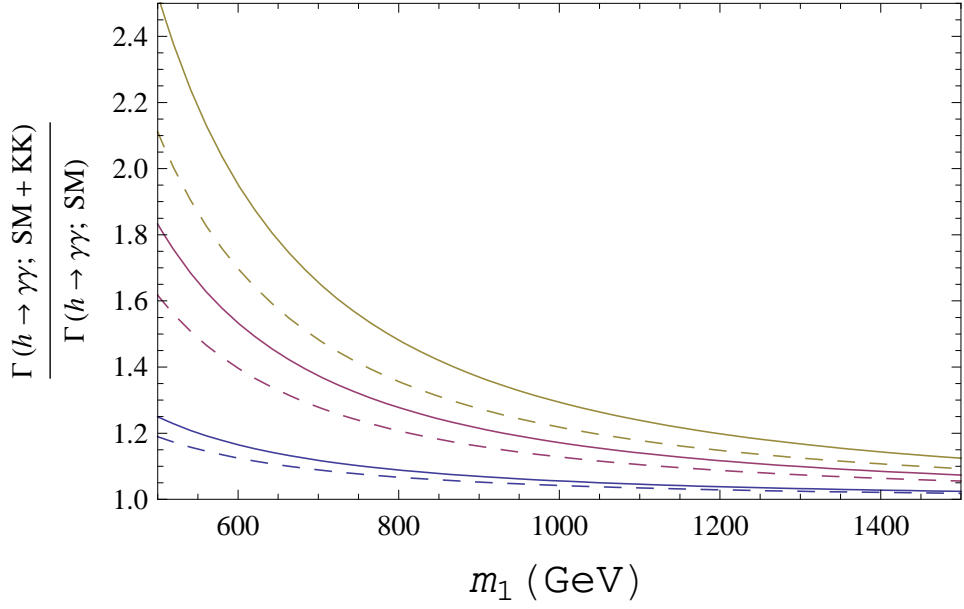}
\caption{
The ratio of the Higgs boson partial decay widths 
 into two photons 
 in the gauge-Higgs unification scenario and in the SM, 
 as a function of the KK mode mass $m_1$. 
The solid lines represent the results 
 including the $n_t$ periodic KK fermion contributions. 
Each solid line corresponds to 
 $n_t=$1, 3 and 5 from bottom to top at $m_1=500$ GeV. 
The results for the $n_t$ half-periodic fermions 
 are depicted as the dashed lines, 
 corresponding to $n_t=$1, 3 and 5 
 from bottom to top at $m_1=500$ GeV. 
}
\label{fig3}
\end{figure}

Next, we analyze the ratio of the partial Higgs boson decay width 
 in the gauge-Higgs unification model to the SM one. 
The KK mode contribution to the effective coupling 
 between Higgs boson and two photons 
 can alter the coupling from the SM one. 
The ratio of the partial Higgs boson decay width 
 into two photons is given as 
\bea 
  \frac{\Gamma(h \to \gamma \gamma;~{\rm SM+KK})}
       {\Gamma(h \to \gamma \gamma;~{\rm KK})}  =
  \left( 1+ \frac{C_\gamma^{KK(GH)}}{C_\gamma^{SM}} \right)^2 . 
\eea
The ratio as a function of $m_1$ 
 is depicted in Fig.~\ref{fig3} 
 for both the periodic and half-periodic fermions 
 and for $n_t=$1, 3 and 5. 
The KK mode contribution is constructive to the SM result.  
For $m_1=1$ TeV and $n_t=1$, 
 the deviation from the SM result is small, about 5\%. 
As $m_1$ is lowered and $n_t$ is raised, 
 the KK mode contributions are dominating as expected. 
The Higgs boson branching ratio into two photons is very small 
 and thus, this ratio can be approximated 
 as the ratio of the partial decay width into two photon 
 in the gauge-Higgs unification model to the SM one. 
This ratio directly reflects the number of two photon events, 
 at the LHC, from the Higgs production through weak-boson fusion 
 and Higgs decay into two photons, when Higgs boson is light.

\begin{figure}[t]
\includegraphics[scale=1.2]{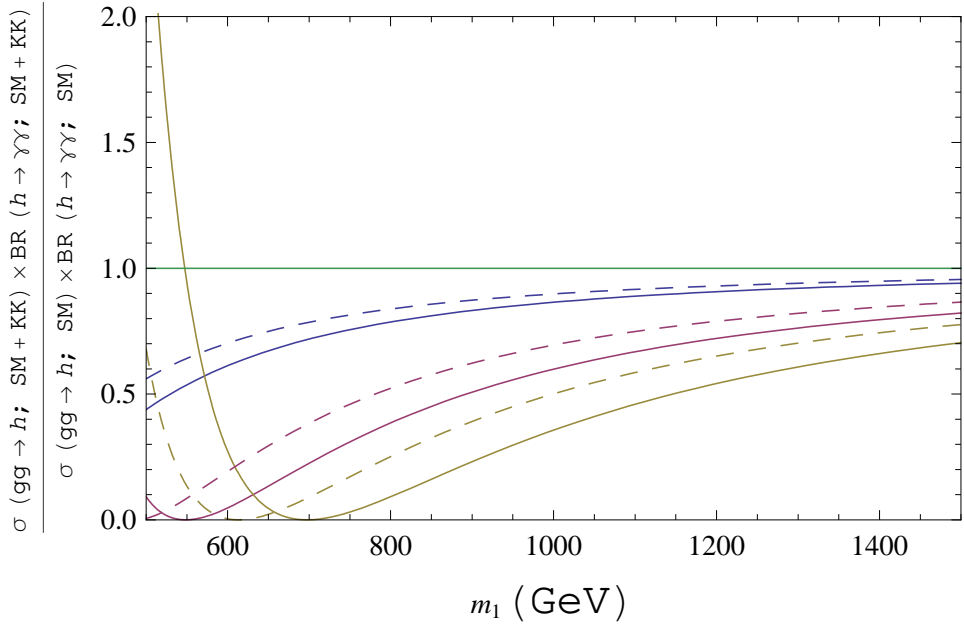}
\caption{
The ratio of the number of two photon events 
 in the gauge-Higgs unification scenario 
 with $n_t$ periodic and half-periodic KK modes 
to those in the SM, 
 as a function of the KK mode mass $m_1$. 
The solid lines represent the results 
 including the periodic KK fermion contributions. 
Each solid line corresponds to 
 $n_t=$1, 3 and 5 from top to bottom at $m_1=1500$ GeV. 
The results for the $n_t$ half-periodic fermions 
 are depicted as the dashed lines, 
 corresponding to $n_t=$1, 3 and 5 
 from top to bottom at $m_1=1500$ GeV. 
}
\label{fig4}
\end{figure}

Finally, we show the ratio of the number of two photon events  
 from Higgs decay produced through gluon fusion at the LHC. 
As a good approximation, 
 this ratio is described as 
\bea 
&& \frac{
   \sigma(gg \to h;~{\rm SM+KK}) 
      \times BR(h \to \gamma \gamma;~{\rm SM+KK})}
   {\sigma(gg \to h;~{\rm SM}) 
      \times BR(h \to \gamma \gamma;~{\rm SM})} \nonumber \\ 
&\simeq & 
\frac{\sigma(gg \to h;~{\rm SM+KK}) 
      \times \Gamma(h \to \gamma \gamma;~{\rm SM+KK})}
   {\sigma(gg \to h;~{\rm SM}) 
  \times \Gamma(h \to \gamma \gamma;~{\rm SM})} \nonumber \\ 
&=& 
 \left(1+ \frac{C_g^{KK(GH)}}{C_g^{SM}} \right)^2 
 \left(1+ \frac{C_\gamma^{KK(GH)}}{C_\gamma^{SM}} \right)^2 .
\eea 
Fig.~\ref{fig4} shows the results 
 for the periodic and half-periodic KK modes 
 as a function of $m_1$ for $n_t=$1, 3 and 5. 
Even for $m_1=1$ TeV and $n_t=1$, 
 the deviation is sizable $\simeq$14\%. 
When $m_1$ is small and $n_t$ is large, 
 the new physics contribution can dominate.

\section{Conclusions and discussions}
In the gauge-Higgs unification scenario, 
 we have discussed the KK mode contributions 
 to the effective couplings between Higgs boson 
 and gauge bosons (gluons and photons). 
Even in the Standard Model, the effective couplings 
 are induced through loop corrections, 
 so that the KK mode contributions can be sizable. 
At the LHC, the main production process of Higgs boson 
 is through gluon fusion and if Higgs boson is light, 
 the primary discovery mode is its two photon decay. 
Therefore, the effects on the effective couplings 
 in the gauge-Higgs unification have a great impact 
 on the Higgs boson search at the LHC. 

We have calculated the fermion KK mode contributions to 
 the Higgs effective couplings through one-loop diagrams 
 and found that the contributions are finite nevertheless 
 the summation is taken for the infinite tower of KK states. 
This finiteness  is achieved by a non-trivial cancellations 
 between two KK mass eigenstates, 
 each of whose contributions is divergent. 
The overall sign of the contributions is opposite 
 compared to the SM result by top quark loop corrections 
 and the similar result in the UED scenario. 
Therefore, this feature is the key to distinguish 
 the gauge-Higgs unification from the UED scenario. 
Our analysis have shown that even with the KK mode mass 
 is around 1 TeV, the KK mode loop corrections provide 
 ${\cal O}$(10\%) deviations from the SM results 
 in Higgs boson phenomenology at the LHC: 
Higgs boson production cross section through gluon fusion 
 is reduced by ${\cal O}$(10\%), 
 the branching ratio into two photon is increased by about 10\%,  
 and the number of two photon evens from Higgs boson 
 is reduced by ${\cal O}$(10\%). 
In a realistic gauge-Higgs unification model, 
 some extra top-like quarks would be introduced  
 to reproduce the top Yukawa coupling in the SM. 
If this is the case, 
 the KK mode contributions are enhanced and can dominate 
 over the SM one. 
In a realistic model, 
 the signal events of Higgs boson production at the LHC 
 are quite different from those in the SM.

Remarkable feature of the KK mode contributions to 
 the effective couplings 
 is that the overall sign is opposite to 
 the results in the SM and the UED scenario. 
Interestingly, the same results (opposite sign) 
 have been found in other models such as 
 the little Higgs model \cite{CTY} and supersymmetric models 
 \cite{Dermisek:2007fi}, 
 all of which are free from the problem of the quadratic divergence 
 in Higgs mass squared corrections (at least, at one-loop level). 
Although we do not have a definite opinion 
 on this opposite sign issue 
 for the time being, this may have something to do 
 with the Higgs mass squared corrections. 
This is because the one-loop diagrams providing 
 the effective couplings between Higgs boson 
 and gluons/photons can be obtained from the one-loop Higgs 
 boson self-energy diagram by attaching two gauge boson external 
 lines and replacing one of the Higgs boson field into its VEV. 
A model which is free from the quadratic divergence 
 of the Higgs self-energy (at one-loop level) may 
 always provide the opposite sign to the effective Higgs coupling. 

Finally, a few comments are in order. 

We have considered the gauge-Higgs unification model in flat space. 
In a simple gauge-Higgs unification model, 
it is known that the lightest KK mode appears around the electroweak scale, 
to realize the electroweak symmetry breaking 
with the correct Higgs VEV in effective Higgs potential. 
For a realistic model, 
we need to generate a hierarchy between the electroweak scale 
and the KK mode mass. 
In an elaborate gauge-Higgs model in flat space (see, for example, \cite{CCP}), 
this situation is realized by introducing several additional bulk fermions 
(and fermions in higher representations). 
As mentioned in the previous section, 
such new KK fermions give additional contributions to effective Higgs couplings. 
In a realistic gauge-Higgs unification model in flat space, 
it would be natural that the KK mode contribution dominates over the SM one.

Recently, the gauge-Higgs unification in the warped background 
has been recently paid much attention, where the hierarchy 
between the electroweak scale and the KK mode mass is realized 
by non-trivial Higgs zero-mode function. 
We expect that our results hold true even in the gauge-Higgs models 
on the warped background. 
Namely, the overall sign of the effective couplings from the KK mode loop 
is opposite to that of the SM and the UED. 
However, note that in the warped case, the nontrivial Higgs zero 
mode function induces a mixing between the SM top quark and 
its KK modes, so that the coupling between Higgs boson and 
the SM top quark is reduced. 
This effect should also be taken into account in the calculations 
for the processes $gg \to h$ and $h \to \gamma \gamma$.

\vspace*{5mm}

{\bf Note added}: 
After completing this work, we were aware of the recent paper 
 by Falkowski \cite{Falkowski:2007hz}, 
 where basically the same subjects are discussed 
 and the similar results are presented. 
Our result presented in this paper is based on 
 the talk given by N. Okada on January 10, 2007
 at a mini workshop held at National Center 
 for Theoretical Sciences (NCTS), 
 National Tsing Hua University, Taiwan. 
We are also aware of that the related subjects have been 
 discussed by a few seminar talks by I. Low 
 (in collaboration with R. Rattazzi) this year \cite{LR}.

\subsection*{Acknowledgment}
The work of authors was supported 
in part by the Grant-in-Aid for Scientific Research 
of the Ministry of Education, Science and Culture, No.18204024 (N.M.) 
and No.18740170 (N.O.).

\newpage

\end{document}